\documentclass{aastex63}

\usepackage{threeparttable}
\shorttitle{Identifying Strong Gravitational Lenses}
\shortauthors{Teimoorinia and Toyonaga}
\begin{document}

\title{Comparison of Multi-Class and Binary Classification Machine Learning Models in Identifying Strong Gravitational Lenses}

\correspondingauthor{Hossen Teimoorinia}
\email{hossen.teimoorinia@nrc-cnrc.gc.ca, hossteim@uvic.ca}

%\author[0000-0002-0786-7307]{Hossen Teimoorinia}
\author{Hossen Teimoorinia}
\affiliation{NRC Herzberg Astronomy and Astrophysics, 5071 West Saanich Road, Victoria, BC, V9E 2E7, Canada }
 \affiliation{Department of Physics and Astronomy, University of Victoria, Victoria, BC, V8P 5C2, Canada}

\author{Robert D. Toyonaga}
\affiliation{NRC Herzberg Astronomy and Astrophysics, 5071 West Saanich Road, Victoria, BC, V9E 2E7, Canada }
 \affiliation{ Department of Electrical and Computer Engineering, University of Waterloo, Waterloo, ON, N2L 3G1, Canada}

\author{ Sebastien Fabbro}
\affiliation{NRC Herzberg Astronomy and Astrophysics, 5071 West Saanich Road, Victoria, BC, V9E 2E7, Canada }
 \affiliation{Department of Physics and Astronomy, University of Victoria, Victoria, BC, V8P 5C2, Canada}

\author{Connor Bottrell}
 \affiliation{Department of Physics and Astronomy, University of Victoria, Victoria, BC, V8P 5C2, Canada}

\begin{abstract}

    Typically, binary classification lens-finding schemes are used to discriminate between lens candidates and non-lenses. However, these models often suffer from substantial false-positive classifications. Such false positives frequently occur due to images containing objects such as crowded sources, galaxies with arms, and also images with a central source and smaller surrounding sources. Therefore, a model might confuse the stated circumstances with an Einstein ring. It has been proposed that by allowing such commonly misclassified image types to constitute their own classes, machine learning models will more easily be able to learn the difference between images that contain real lenses, and images that contain lens imposters. Using Hubble Space Telescope (HST) images, in the F814W filter, we compare the usage of binary and multi-class classification models applied to the lens finding task. From our findings, we conclude there is not a significant benefit to using the multi-class model over a binary model. We will also present the results of a simple lens search using a multi-class machine learning model, and potential new lens candidates.

\end{abstract}

\keywords{cosmology: observations -- gravitational lensing: strong -- methods: data analysis  -- methods: numerical -- methods: observational -- techniques: image processing}

\section{Introduction}
\label{introduction}

The principles of general relativity predict that the path of a light ray will be deflected in the presence of a massive object (i.e., a  gravitational lens) such as a galaxy or cluster of galaxies \citep{Einstein36}. In this situation, the massive object serves as a lens. When the lens is located between an observer and a source, depending on the shape of the source,  different phenomena can be observed. A point source may be observed as multiple sources \citep{Walsh79}, whereas an extended background source may be detected as arcs or rings (i.e., the Einstein ring). These phenomena are significant because,  for example, the multiple images of a background point source can follow different paths to the observer and can generate a time-delay between those images that may be used to constrain cosmological parameters, such as the Hubble constant \citep{Kundic97}. The case in which a massive lensing object generates multiple images of a background source is called strong gravitational lensing. Cases of strong lensing are relatively rare because a proper alignment of a source, the lens and the observer is needed to create the effect. In weak lensing multiple images of a background source are not created. Instead, a systematic alignment of the shapes of background sources around the lensing mass is created \citep{Tyson90}. In other words, a statistical analysis, rather than a single strong effect, is the matter of the investigation. Gravitational lensing overall is one of the most powerful tools for examining the matter content and the overall geometry of the Universe and also for studying distant galaxies \citep{Treu10}.

In searching for strong gravitational lenses, traditional techniques typically rely on methods that discriminate based on features such as the presence of arcs in images, \citep{alard06,more12} or that are based on spectroscopic information \citep{faure08}. Such methods often return high false-positive rates, so they also require much human analysis. Additionally, these methods lack generality. For example, arc-finding algorithms generally have difficulty classifying lenses that either do not have prominent Einstein rings or else have a morphology that does not closely approximate an arc \citep{jacobs19}. For this reason, such methods are generally primarily effective at identifying specific types of gravitational lenses.

Machine learning-based approaches have begun to be used as an efficient and effective method to search for strong gravitational lenses in image data. The most common machine learning approach uses a convolutional neural network model to identify different objects in images. Convolutional neural networks have recently been applied to lens-finding in the Kilo Degree Survey \citep{petrillo17,khramtsov19}, the COSMOS field in HST data \citep{pourrahmani18}, and to CFHTLS data \citep{jacobs17}. These models learn to understand image data and are commonly used to solve computer vision problems. Such methods are effective, but these  models still falsely classify lens imposters that contain features that might be interpreted as a lens – but are not.

Images containing multiple sources are often misclassified as lenses. A possible explanation for such misclassification is that multiple sources occasionally align themselves so as to form what the machine learning model has learned to identify as gravitational lensed source image morphology. Jacobs et al. (2019) propose that considering a third ``crowded sources" class (in addition to images containing lenses and images containing non-lenses with a single source) might help to decrease the misclassification rate. In this paper, we explore the use of a three-class classification scheme and compare the results with binary classification. 

In this paper we will keep specific technical components, such as convolutional neural network (CNN) architectures, to a minimum and focus on the results. For more technical discussions, we refer readers to the papers mentioned above. In section \ref{sec:data}, we describe the data used in this paper, and in section \ref{sec:method} we explain our method and also introduce the metrics we use to compare the two classification models. The results and the comparisons are presented in section \ref{sec:results} as well as a result of searching for real lenses in images. We present the conclusion in section \ref{sec:discusion}.

\section{Data}
\label{sec:data}

\begin{figure*}
% \begin{minipage}{\linewidth}
  \centering
  \begin{tabular}{cccc}

  \includegraphics[width=4cm,height=4cm]{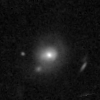}
    & \includegraphics[width=4cm,height=4cm]{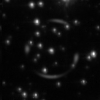}
    & \includegraphics[width=4cm,height=4cm]{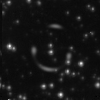}
    & \includegraphics[width=4cm,height=4cm]{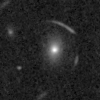}\\
    
   \includegraphics[width=40mm,height=40mm]{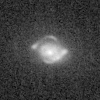}
    & \includegraphics[width=4cm,height=4cm]{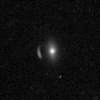}
    & \includegraphics[width=4cm,height=4cm]{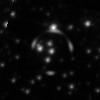}
    & \includegraphics[width=4cm,height=4cm]{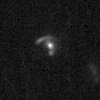}\\
 \end{tabular}
\caption{Examples of 6 simulated gravitational lenses. Logarithmic scaling has been applied to enhance lens features.}
\label{fig:sixlenses}
\end{figure*}

\label{sec:data} 

In a machine learning problem, gathering a suitable training set is an essential step. To provide a suitably sized training set, we use HST images taken with the F814W filter as well as the software LensTool \citep{jullo07} to simulate gravitational lensing effects. In this section, we explain in further detail how we created these datasets.

\subsection{Defining the Crowded Class}

The classes we have chosen to be represented in the multi-class model are lens, single-source non-lens, and crowded source non-lens. After exploring the data using simple CNN models and examining the false positives returned by them, it was observed that images with greater than approximately ten visible sources and the absence of a single dominant source constitute what we would label the ``crowded" class. To test this definition, we selected approximately 1000 images that met our crowded class definition and generated a confusion matrix to determine if a model could distinguish images that met our definition of ``crowded" from lens and traditional non-lens images (Figure \ref{confusion-matrix}). A more detailed description of our class selection procedure is found in section 3.2. 
\subsection{Creating Simulated Images of Gravitational Lenses}
\label{sec:simualtion} 

One obstacle to using machine learning methods, including a convolutional neural network-based approach, is that a substantial set of training data is required to train models. The number of training examples depends on the nature of the data under study, but can typically be on the order of $10000$ examples \citep{lecun98,Bottrell19}. However, since there are only a few hundred unique confirmed lenses known to science, limiting training examples to real lens images is not the most effective approach to training a convolutional neural network. A common method of solving the problem of limited real training data is to simulate gravitational lenses using ray-tracing software \citep{jacobs17,  petrillo17,pourrahmani18}. We adopt a similar approach to that used by \cite{pourrahmani18}. The ray-tracing software that we use is Lenstool, which was developed by the Laboratoire d'Astrophysique de Marseille \citep{jullo07}. This software was also used by Pourrahmani et al. (2018). First, we created simulated lensed source image features, using LensTool code. Then, the simulated sources were combined with a set of cut-out images to create the training set. We will describe these steps in detail. The Python package AstroPy is a commonly used software package in the field of astronomy and was useful in handling FITS image files \footnote{https://www.astropy.org/}. 

First, we used the GOODS-MUSIC catalogue \citep{santini09}, which provides spectroscopic or photometric redshifts for many galaxies. We used these galaxies as deflectors. To cut the galaxies out of HST images, we utilized the Canadian Astronomy Data Centre (CADC) direct data service \footnote{http://www.cadc-ccda.hia-iha.nrc-cnrc.gc.ca/en/doc/data/} and created images of 100$\times$100 pixels. Then Source Extractor \citep{bertin96} was used to verify that a source above $3\sigma$ brighter than the background was present in the centre of the image. This step was necessary because sometimes the source described in the catalogue was either too faint in the downloaded image, or off-centre. 

To make deflector cut-outs of crowded sources, Source Extractor was used to make cut-outs of random HST ACS/WFC images taken with the F814W filter from the CADC database. These images were manually inspected until ~1000 100x100 pixel cut-outs of crowded source images were obtained.

When creating simulated lens images, we are primarily concerned with the morphology of the image of the lensed source. Although a more focused selection of deflector is possible, we opt to indiscriminately select deflectors. Instead of searching for new lenses, our primary goal is to investigate the differences between using binary and multi-class models. Leaving our selection of deflector unrestricted increases the size of our training set and can allow us to compare how well each model type can identify lensing effects alone, independent of the deflector. 
Additionally, by simply introducing the crowded class, models would probably learn to immediately label images with multiple sources as non-lenses without actually searching for the presence of images of the lensed source. To avoid this, we also superimposed lensing features onto the cut-out images of the crowded class and grouped them into the positive training data class. This further highlights the focus on features of lensed source image morphology and less on the choice of deflector.

After the deflector cut-outs were made, simulated lens features were created using LensTool. For the single source deflectors, we utilized the redshifts provided in the GOODS-MUSIC catalogue as input to the LensTool software. First, we used spectroscopic redshifts, or if these were not available, photometric redshift values. Other parameters used by LensTool such as alignment of the lensed source and deflector, magnitude, ellipticity, and rotation were randomized. The redshift of the lensed source was also randomized with LensTool so that it had to be larger than the redshift of the corresponding deflector provided in the GOODS-MUSIC catalogue. Randomization allowed a broader number of morphologies and image possibilities to be represented. With a large number of diverse training examples, CNN models should be able to detect the essential characteristics of a gravitational lens. Unlike the single-source deflectors selected from the GOODS-MUSIC catalogue, redshift information was not available for the crowded-source deflectors, so this parameter was randomized as well.

The final step in creating simulated gravitational lenses was merging images containing lens features with the corresponding images from deflector. A combined image was obtained according to the following equation.\\

$$\rm{combined = S(deflector + \alpha*S( C( lens ) ) )}$$

\begin{itemize}
  \item combined: Image of merged deflector cut-out and lens features 
  \item deflector: The deflector image cut-out
  \item lens: The image containing the simulated lens features
  \item  S(): A function that scales values to a range of [0,1]
  \item $\alpha$: Defined as the maximum pixel value of the deflector cut-out image 
   \item C(): A function that performs kernel convolution using a Gaussian kernel with FWHM =0.986 pixels to mimic PSF of  F814W images.
\end{itemize}

In Fig. \ref{fig:sixlenses} we show an example of 6 simulated gravitational lenses created by LensTool code, to which different sources were added according to the method described above. 

\subsection{ The Non-Lens Classes and Data Augmentation}
The deflector image cut-outs for single source non-lens and crowded source non-lens classes were made similarly to the cut-outs for the gravitational lens class. Instead of being merged with an image of lensing features, the pixel values of these images were normalized between zero and one. Then, each of the images in the three different classes underwent data augmentation to increase the number of available images. The transformations that were performed were: reflections over the vertical and horizontal axis, 90$^{\circ}$  rotations, and random rotations within the range of [1$^{\circ}$ ,89$^{\circ}$ ] and [-1$^{\circ}$ ,-89$^{\circ}$ ].  This resulted in approximately ~12000 images to be used for training from each of the three classes of images. Data augmentation is a technique that increases the completeness of the dataset with the goal of helping machine learning models better generalize their understanding of image classes.  In our case, reflecting and rotating the images in our dataset helps the models dissociate their understanding of the classes of objects from a specific orientation in a 100x100 pixel image. For example, if four slightly rotated images of the same lens exist in the training set, a model will learn to place emphasis on the morphology of the lens, not on its orientation.

\section{THE METHOD}
\label{sec:method}
In this section, we first summarize the CNN architectures which will be used to assess whether a multi-class or binary classifier is better equipped for lens detection. Second, we address the question of how to choose suitable classes for the multi-class model. Lastly, we describe the training and evaluation of our CNN models.

\subsection{Classifier Model Architecture}
\begin{figure}
\centering
\includegraphics[width=8.cm,height=15cm]{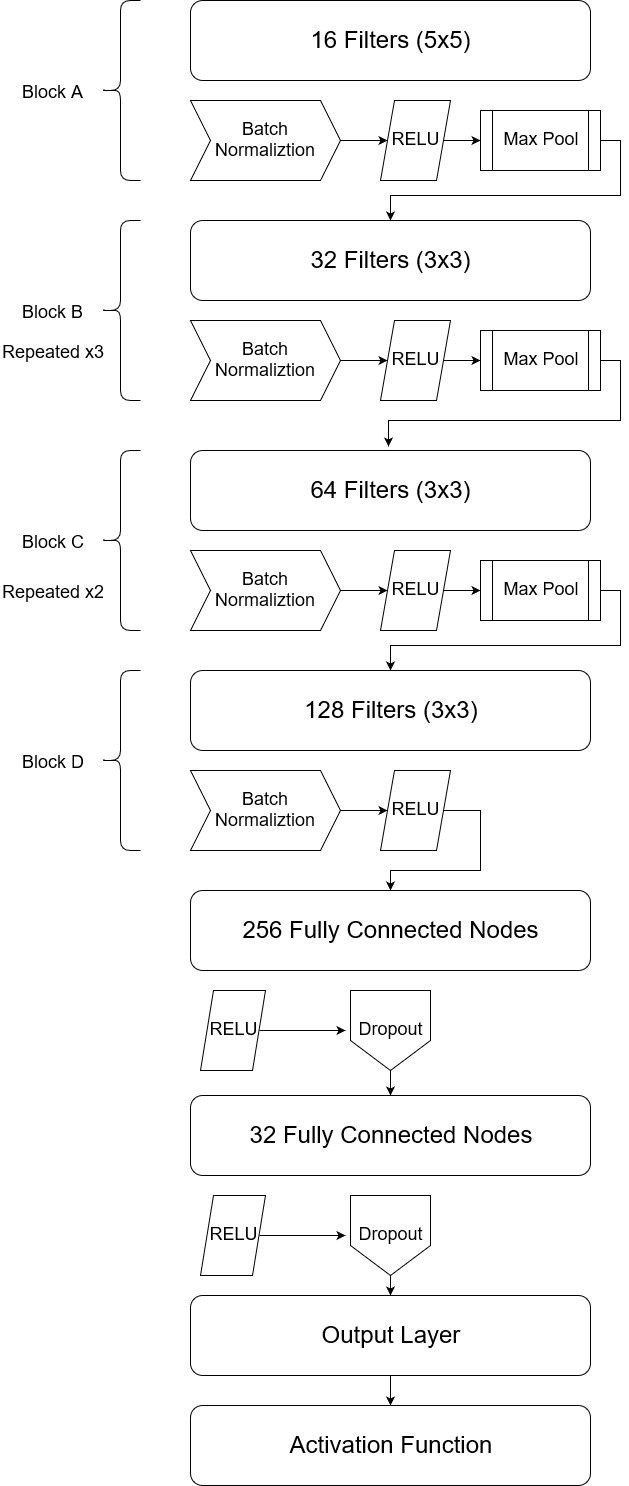}
\caption{A diagram of the CNN model architecture. The first and last block of section B employ max pooling. Also, the last block of section C employs max pooling.}
\label{fig:model}
\end{figure}

The network architecture of both the binary and multi-class models were identical except for specific attributes that allowed the multi-class model to perform predictions in three classes instead of two. Both models had nine convolution layers. Each layer was followed by a batch-normalization step and a rectified linear activation function. Four max-pooling steps were also included which reduced input size from 100x100 to 6x6. As can be seen in Fig. \ref{fig:model},  after the convolution layers, the inputs passed through 3 fully connected layers and then an activation function. In the case of the binary model, the activation function is a sigmoid function. In the case of the multi-class model, the activation function is a softmax function. Both multi-class and binary classification models utilise an Adam optimizer. The multi-class model uses a categorical cross entropy loss function, while the binary model uses a binary cross entropy function. The Tensorflow library \footnote{https://www.tensorflow.org/} is a very common machine learning framework and was used in this work to construct our CNN models.

\subsection{Choosing suitable classes}

We chose the ``crowded" class as the third class to be included in the multi-class classification scheme. The new class was chosen because the identification of crowded sources was a common source of false positives in similar works (Jacobs et al. 2019). We also performed a preliminary test using a binary classifier and trained only on examples with deflector which are found in the GOODS-MUSIC catalogue. Since this test model was primarily trained on images with single sources, we found that, when applying this model, the majority of false positives were images containing crowded sources. This test shows that crowded sources do cause confusion for the CNN model and should be better represented in the training set.

Initially, four classes were chosen to be a part of the multi-class classification model scheme. The fourth ``multisource" class consisted of images containing a dominant source and a small number of surrounding sources – fewer than ten. The surrounding sources might be falsely interpreted as lens features. The difference between these images and those in the ``crowded" class was that the ``crowded" class had more than ten sources contained in the image and need not have a central dominant source. 

\begin{figure}
\centering
\plotone{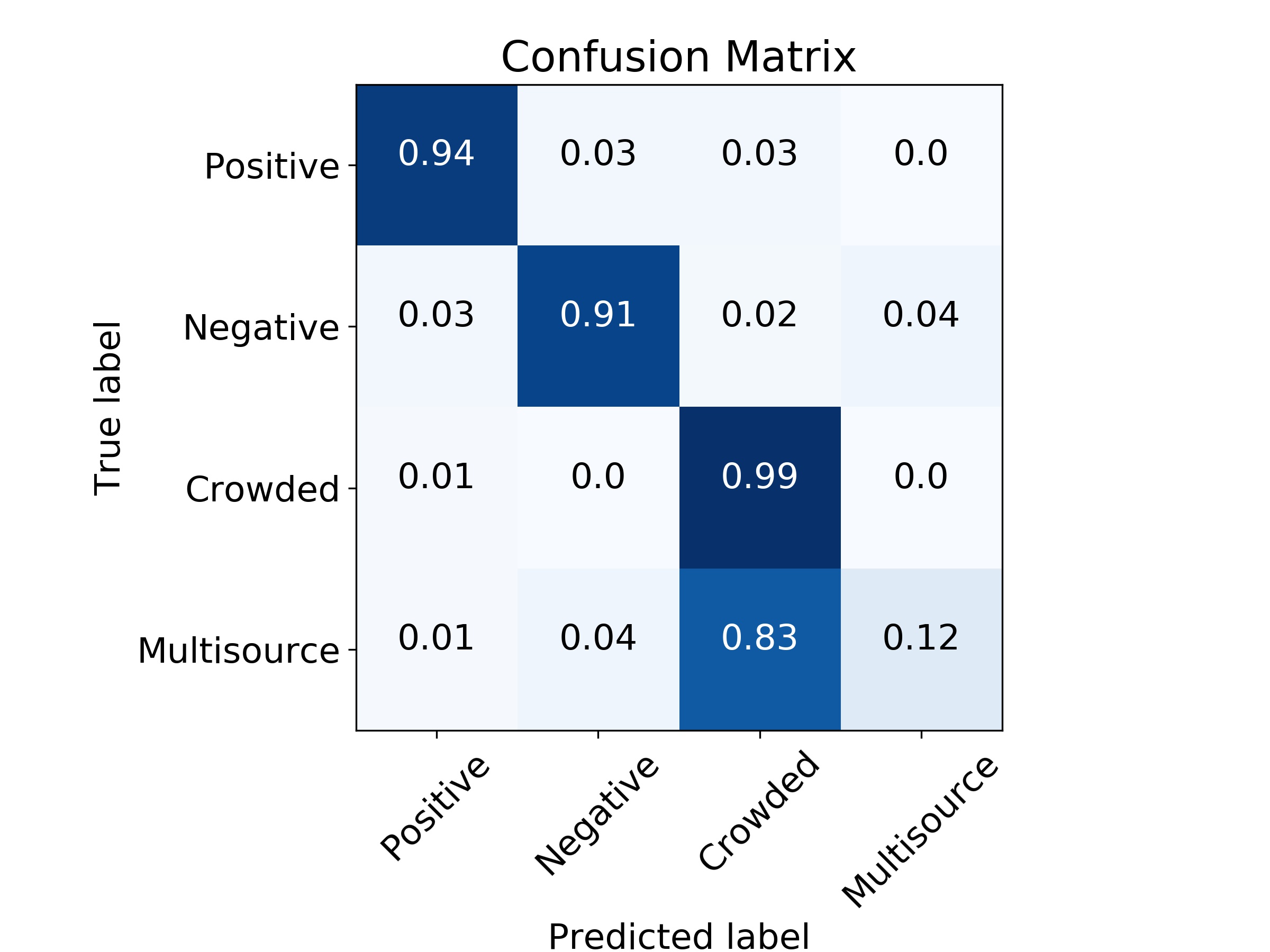}
\caption{The plot shows a (percentage) confusion matrix with four classes. The crowded class is not easily distinguishable from the fourth ``multisource" class. }
\label{confusion-matrix}
\end{figure}

To test whether the four classes (lens, single-source negative, crowded source negative, fourth class negative) were distinct, we trained a 4-class classification model on the aforementioned image types. Then, we tested the model on $\sim2000$ test images from each class to generate a confusion matrix, as seen in Figure \ref{confusion-matrix}. We can see from the confusion matrix that the fourth class is easily confused with images of the ``crowded" class. The lower matrix indicates that the fourth class was not distinct and should not be included in the multi-class classification scheme as its own individual class. 

\subsection{The metrics}
\label{sec:metric}

We will compare the performance of the two models in this paper using various metrics. 
In a classification problem, accuracy measures the ratio of the correct predictions on a sample, that is:

\begin{equation}
  accuracy=\frac{TP +TN}{FN+FP+TP+TN} 
  \label{eq:Bsp_OhmsLaw}
\end{equation}

\begin{equation}
  precision=\frac{TP}{TP+FP} 
  \label{eq:Bsp_OhmsLaw}
\end{equation}

\begin{equation}
  recall=\frac{TP}{TP+FN} 
  \label{eq:Bsp_OhmsLaw}
\end{equation}

\begin{equation}
  F1=2\times\frac{(Precision)(Recall)}{Precision + Recall} 
  \label{eq:Bsp_OhmsLaw}
\end{equation}

TP (True Positive) is the number of positive instances that were correctly predicted. TN (True Negative) is the number of negative cases that were correctly predicted (here, the Negative class includes all the other classes). Moreover, FN (False Negative) indicates the number of instances that belong to the positive group but were classified as Negative. However, FP (False Positive) shows the number of examples that belonged to the negative group but were predicted as positive.

A high value for precision shows the power of an algorithm to minimize the number of examples incorrectly identified as FP while recall describes the ability of the algorithm to minimize the number of occurrences wrongly identified as FN. A high  F1 score implies that both precision and recall are both high, which is a representation of a perfect prediction.

\subsection{Model Training and Evaluation}
\label{sec:evaluation}
We independently trained ten binary models and ten multi-class models, using 12000 training examples per class. In the case of the multi-class model, this means that 12000 images of lenses, 12000 images of single-source negatives, and 12000 images of crowded negatives were included in the training set. In the case of the binary model, 6000 images of single-source negatives and 6000 images of crowded source negatives were concatenated to form 12000 images in the ``negative" class. The same 12000 images of lenses were used for the binary ``positive" class. It should be observed that this scheme results in the multi-class model learning from a total of 36000 examples and the binary model learning from 24000.  We believe this is an appropriate scheme because it keeps the number of training examples per class consistent (there is no class imbalance in either model's training set). For completeness, we explore other training set configurations as well. In alternate configuration 1, we maintain the same examples for the binary model as before, but select 12000 positives, 6000 single source negatives, and 6000 crowded negatives for the multi-class training set. This results in both models training from 24000 samples, but now there is an imbalance in the multi-class model's training set.  In alternate configuration 2, we maintain the same examples for the multi-class model as before, but select 12000 positives, 12000 single source negatives, and 12000 crowded negatives for the binary model's training set. This results in both models training from 36000 samples, but now there is an imbalance in the binary model's training set.

First, we trained each of the models. Then, to evaluate the performance of the models, we used a balanced validation set of the total number of 7500 images, which were not part of the training set. Training continued until the lowest amount of validation loss was reached for each model. We used the lowest loss as our metric for when to stop training instead of using a fixed number of epochs, because each model may take a different number of iterations to train, due to random initialization of trainable parameters.

After training the models, we evaluate them by gathering accuracy, precision, recall, and F1 score performance metrics. To do this, we use the average predictions of ten binary and ten multi-class models on our test set. We also calculate a 95 \% confidence interval using the ten models of each variety. The test set consists of 2500 lens images, 2500 single source negatives, and 2500 crowded source negatives. To assess the performance at different decision boundaries, we also generate a ROC plot and AUC scores for both models using a set of approximately 8500 negative (single or crowded sources) samples randomly chosen from HST images and 3000 simulated positive examples.
\section{Results}
\label{sec:results}

In this section we compare the performance of a multi-class classifier to a binary classifier. To demonstrate the applicability of such models to a practical lens search we present the results of a rudimentary lens search on HST images. Specifically, we present some known lenses that we were able to recover, as well as potential new lens candidates that we were unable to find in previous literature.

\subsection{Comparison of Binary to Multi-Class Model Performance}

\begin{figure} % bar graph (original)
\centering
% \plotone{bar-graphs/comparison-alternative1}
% \plotone{bar-graphs/comparison-alternative2}
\includegraphics[scale=0.7]{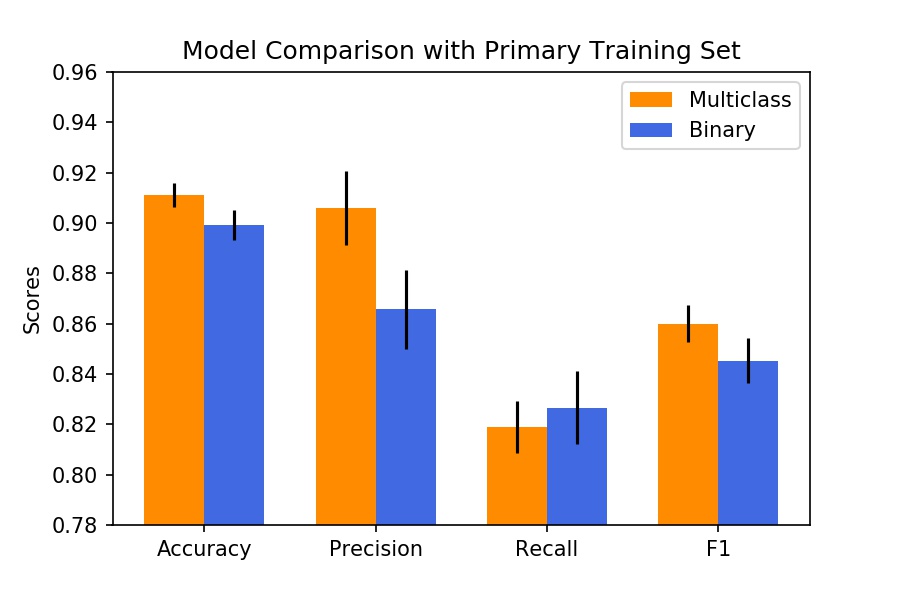}
\includegraphics[scale=0.7]{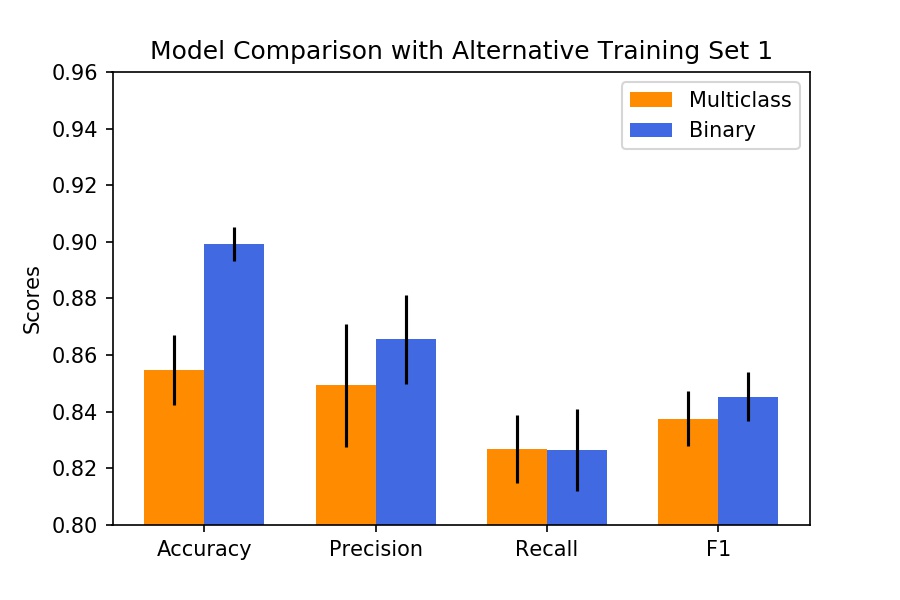}
\includegraphics[scale=0.7]{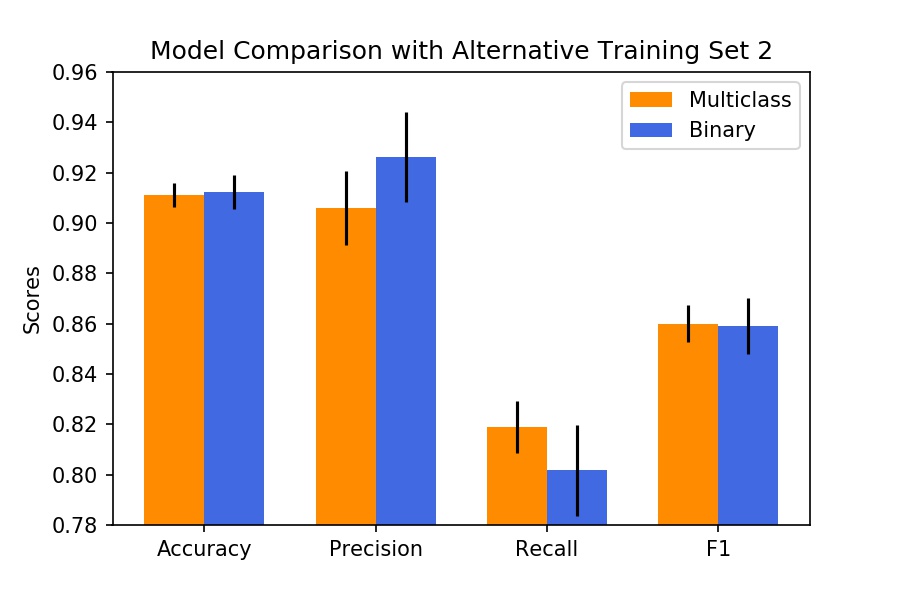}
\caption{These plots show the four different metrics used in this paper to compare the performance of the binary and multi-class models.  Primary training set: 12000 training examples per class.  In the case of the multi-class model, this means 12000 images of lenses, 12000 images of single-source negatives, and 12000 images of crowded negatives. In the case of the binary model, 6000 images
of single-source negatives and 6000 images of crowded source negatives were concatenated to form 12000 images in the ``negative" class. Alternate configuration  1:  we  maintain  the  same  examples  for  the  binary  model  as  before,  but  select  12000
positives,  6000  single  source  negatives,  and  6000  crowded  negatives  for  the  multi-class  training  set. Alternate configuration 2: we maintain the same examples for the multi-class
model as before, but select 12000 positives, 12000 single source negatives, and 12000 crowded negatives
for the binary model’s training set. }
\label{fig-bars}
\end{figure}

We compare the binary and multi-class models using a variety of metrics. The results of these evaluations can be found in Table \ref{metric-table0}. Figure \ref{fig-bars} is provided for a more visual representation of the data. As discussed in Sec. \ref{sec:evaluation}, we investigate alternative training set configurations in addition to our primary investigation. These results of these alternative evaluations are presented in Table \ref{metric-table1} and Table \ref{metric-table2} respectively. Precision or recall alone is not enough to quantify a model's performance because a model can return very low false-positive rates, but also classify nearly all predictions as negative, being unable to identify positives correctly. The converse is also a possibility. For this reason, the F1 score of each of the models is a good way to gauge its performance because it is a function of both precision and recall. From the results of these tests, we cannot conclusively say that the multi-class model is generally more performant than the binary model.

\begin{figure*} 
\centering
\includegraphics[width=18.cm,height=10cm,angle=0]{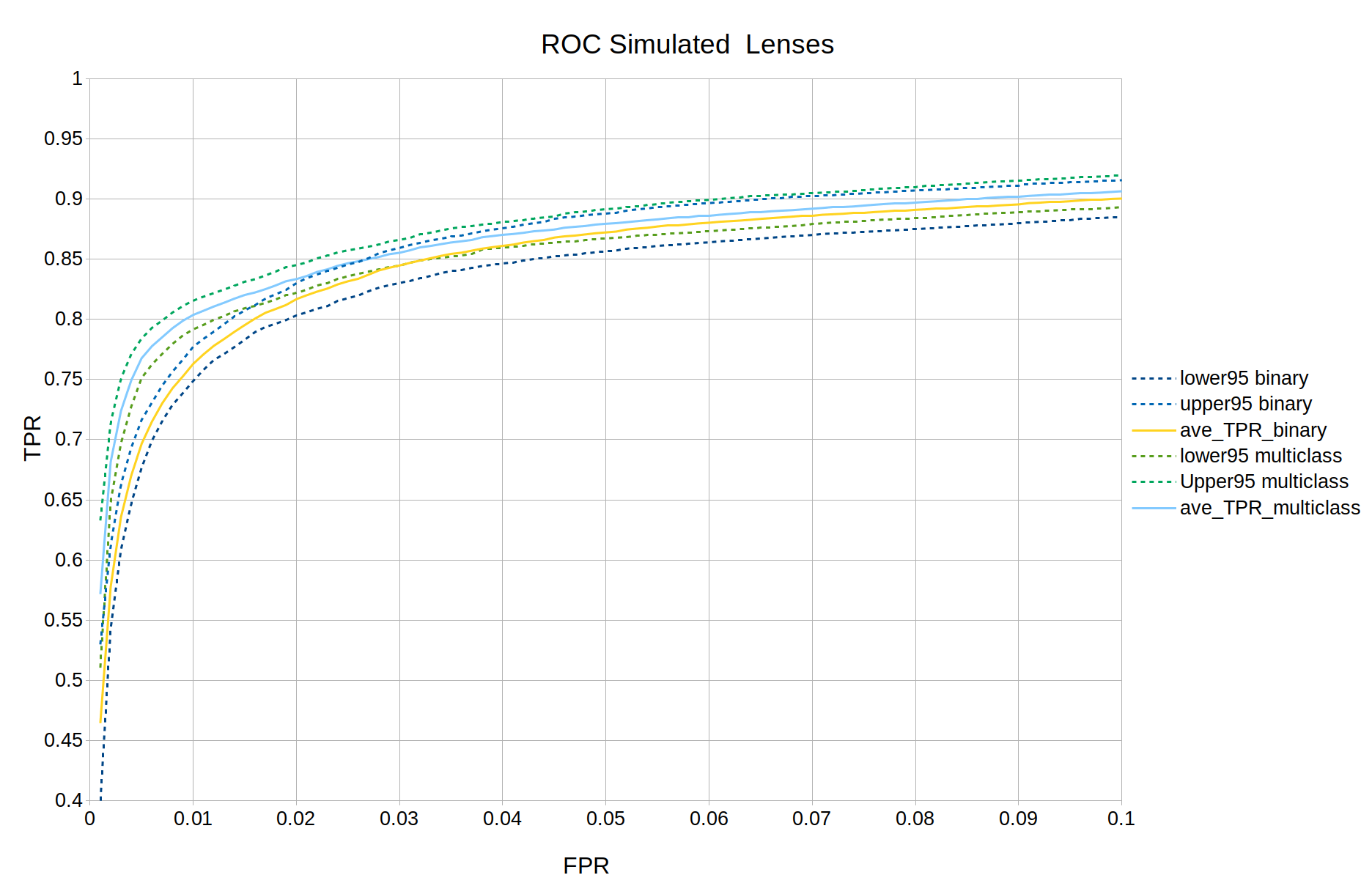}
\caption{ROC Plot: a comparison between the binary and multi-class models.}
\label{ROC}
\end{figure*}

Although the multi-class model appears to out-perform the binary model when trained on our original test set, its relative performance drops when our alternative training sets are used (motivations for which are in \ref{sec:evaluation}). This is an interesting result because it appears that the multi-class model is better when the training sets for both models are balanced (equal numbers of training samples in each class). However, when maintaining a common total number of training examples between models is prioritized over balanced training data, the multi-class model is no longer better. It is general practice, when training machine learning classifiers, to maintain class balance so that models have enough data to learn the discriminating features of each class equally.
However, the argument can be made that since more training images were given to the multi-class model in our original scheme, the binary model was at a disadvantage. It can also be observed that the receiver operating characteristic (ROC) plot in Figure \ref{ROC} shows that both models perform similarly when not restricted to the natural decision boundary of 0.5. ROC plots are often used to characterize a model's performance. They display the degree of recovery of true positive samples with respect to a variable false positive rate. Using the plot, we can generate an area under the curve score (AUC) for each model. A large AUC score means that many true positives can be recovered at low false positive rates. A small AUC score means that high false positive rates are required to achieve high true positive recovery.  AUC ranges between 0.5 and 1. The value of AUC close to 0.5 and 1 shows a random and perfect classification, respectively \citep[e.g., for more information about ROC and AUC plots see ][and references therein ]{Teim16}.  The binary model has an AUC score of  $0.95341 \pm 0.006033\%$ , and the multi class model has an AUC score of $0.95838 +\pm 0.005966$ with a 95\% confidence interval. These scores are not significantly different.

%Table of metric results (original)
\begin{table}
  \centering
 %\begin{threeparttable}
 
\caption{Model Performance Comparison (Primary Training Set)}
\begin{tabular}{ lll }%|p{2.5cm}|p{2.5cm}|p{2.5cm}
\hline Metric & Multi-class Model & Binary Model \\
\hline
%AUC Score    & 0.95838 $\pm$  0.00596 & 0.95341  $\pm$  0.00603\\
Accuracy    & 0.91113 $\pm$ 0.00490 & 0.89920 $\pm$ 0.00589 \\
Precision  & 0.90598 $\pm$ 0.01462 & 0.86568 $\pm$ 0.01571 \\
Recall & 0.81892 $\pm$ 0.01048 & 0.82657 $\pm$ 0.01447 \\
F1 Score & 0.86004 $\pm$ 0.00720 & 0.84536 $\pm$ 0.00877
%\hline
\label{metric-table0}
\end{tabular}

 \hfill \break
 
   \caption{Model Performance Comparison (Alternative Training Set 1)}
    \begin{tabular}{ lll }%|p{2.5cm}|p{2.5cm}|p{2.5cm}
\hline Metric & Multi-class Model & Binary Model \\
\hline
%AUC Score    & 0.95838 $\pm$  0.00596 & 0.95341  $\pm$  0.00603\\
Accuracy    & 0.85481 $\pm$ 0.01228 & 0.89920 $\pm$ 0.00589 \\
Precision  & 0.84934 $\pm$ 0.02169 & 0.86568 $\pm$ 0.01571 \\
Recall & 0.82688 $\pm$ 0.01195 & 0.82657 $\pm$ 0.01447 \\
F1 Score & 0.83753 $\pm$ 0.00974 & 0.84536 $\pm$ 0.00877
%\hline
\label{metric-table1}
\end{tabular}

\hfill \break

  \caption{Model Performance Comparison (Alternative Training Set 2 )}
    \begin{tabular}{ lll }%|p{2.5cm}|p{2.5cm}|p{2.5cm}
\hline Metric & Multi-class Model & Binary Model \\
\hline
%AUC Score    & 0.95838 $\pm$  0.00596 & 0.95341  $\pm$  0.00603\\
Accuracy    & 0.91113 $\pm$ 0.00490 & 0.91241 $\pm$ 0.00666 \\
Precision  & 0.90598 $\pm$ 0.01462 & 0.92628 $\pm$ 0.01779 \\
Recall & 0.81892 $\pm$ 0.01048 & 0.80180 $\pm$ 0.01812 \\
F1 Score & 0.86004 $\pm$ 0.00720 & 0.85913 $\pm$ 0.01116
%\hline
\label{metric-table2}

\end{tabular}

\begin{tablenotes}
     \small
     \centering
     \item The error for each measurement is calculated at 95\% confidence. 

    \end{tablenotes}
% \end{threeparttable}

%\caption{The error for each measurement is calculated at 95\% confidence}
\end{table}

\subsection{Searching for Real Lenses in Images}

\begin{figure*}
% \begin{minipage}{\linewidth}
  \centering
  \begin{tabular}{cccc}

  \includegraphics[width=40mm,height=40mm]{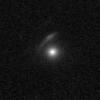}
    & \includegraphics[width=4cm,height=4cm]{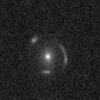}
    & \includegraphics[width=4cm,height=4cm]{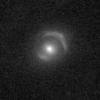}
    & \includegraphics[width=4cm,height=4cm]{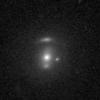}\\
  (342.18204, -44.54032)  & (149.87897, 2.57443) & (179.93597,-0.12451) & (199.62253, -1.07255)\\

  \includegraphics[width=4cm,height=4cm]{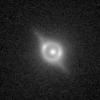} 
    & \includegraphics[width=4cm,height=4cm]{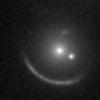}
    & \includegraphics[width=4cm,height=4cm]{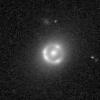}
    & \includegraphics[width=4cm,height=4cm]{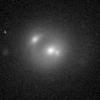}\\
 (221.62575,38.94905) & (204.46373, 36.33838) & (206.77074, -1.01766) & (241.34712, 38.19832) \\  
 
   \includegraphics[width=4cm,height=4cm]{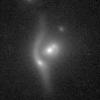}
    & \includegraphics[width=4cm,height=4cm]{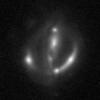}
    & \includegraphics[width=4cm,height=4cm]{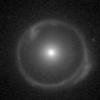}
    &  \includegraphics[width=4cm,height=4cm]{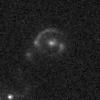} 
   \\
  (241.52964, 22.58644) & (242.30787, 65.54100) & (247.95978, 18.90117) & (219.26350, 35.03143)\\  
  
\\

  \includegraphics[width=4cm,height=4cm]{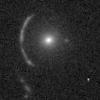}

& \includegraphics[width=4cm,height=4cm]{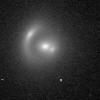}
\\
(34.40471, -5.22489)& (184.52757,56.80131) \\  
  
%   \includegraphics[width=4cm\linewidth,height=4cm]{findings/candidtate_B_197-847108113_-1-347914274.jpg}
%   & (197.84711, -1.34791) 
% & \includegraphics[width=4cm\linewidth,height=4cm]{findings/candidtate_B_177-391763686_22-3905413776.jpg}
% & (177.39176, 22.39054) 
% & \includegraphics[width=4cm\linewidth,height=4cm]{findings/candidtate_A_35-273747_35-937809.jpg}
%  & (35.27375, 35.93781) 

\end{tabular}
\caption{Known lenses recovered through search. (Right Ascension, Declination)}
\label{fig:candidate1}
\end{figure*}

\begin{figure*}
% \begin{minipage}{\linewidth}
  \centering
  \begin{tabular}{cccc}

  \includegraphics[width=40mm,height=40mm]{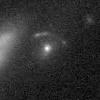}
    & \includegraphics[width=4cm,height=4cm]{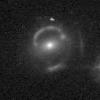}
    & \includegraphics[width=4cm,height=4cm]{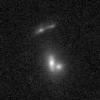}
    & \includegraphics[width=4cm,height=4cm]{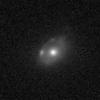}\\
  (245.34260, 38.16843) & (52.42005, -2.22174) & (251.26625,1.66838) & (41.36174, -53.06985) \\  
  
  \includegraphics[width=4cm,height=4cm]{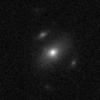}
& \includegraphics[width=4cm,height=4cm]{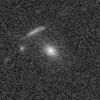}
& \includegraphics[width=4cm,height=4cm]{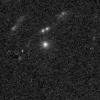}
&\includegraphics[width=4cm,height=4cm]{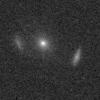}\\
(278.96005, 52.62944) & (28.32948, -13.87092) & (334.45428, 0.37398) & (202.774884291, -41.20207)\\  

   \includegraphics[width=4cm,height=4cm]{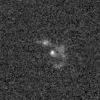}
 & \includegraphics[width=4cm,height=4cm]{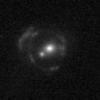}
  & \includegraphics[width=4cm,height=4cm]{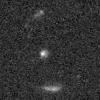} 
  & \includegraphics[width=4cm,height=4cm]{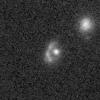}\\
(189.51785, 62.24911)  & (258.31187, 60.36893) & (127.74968, 51.28268)&  (34.42320, -5.21026)\\

\includegraphics[width=4cm,height=4cm]{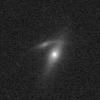}
& \includegraphics[width=4cm,height=4cm]{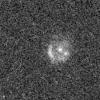}
& \includegraphics[width=4cm,height=4cm]{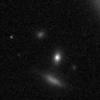}
& \includegraphics[width=4cm,height=4cm]{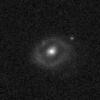}
\\
(147.44679, 17.12524) & (140.30041, 45.45082) & (158.66508, 51.90089) & (219.36531, 34.23377) \\  

% & \includegraphics[width=4cm\linewidth,height=4cm]{findings/candidtate_C_177-39176465_22-3905310969.jpg}
% (177.39176, 22.39053) not sure if found

 \end{tabular}
\caption{Potential lens candidates we were unable to find in previous work. (Right Ascension, Declination)}
\label{fig:candidate2}
\end{figure*}

This work aims to show possible differences in a multi-class classification with a binary model. However, we also show that the models can be applied to a practical lens finding task by performing a rudimentary search on HST images. To that end, we used the multi-class classification model to search HST images for real gravitational lenses. All HST images accessible via the CADC direct data service taken with the ACS/WFC instrument and filter F814W were searched – approximately 10000 full images, resulting in 20 million cutouts. Source Extractor was run on each of the images, and 100x100 pixel cut-outs were created for each of the identified sources. The image cut-outs are then normalized according to the procedure in Section \ref{sec:simualtion} . The image cut-outs were then sent to an ensemble of 3 multi-class CNN models to get a statistical result. Each of the models performed predictions on each of the image cut-outs. The average prediction score of the three models was the resulting output prediction score for each of the image cut-outs. Approximately 20 million predictions were made. We manually inspected the first ~20,000 image cut-outs that received a prediction of over 66.7\% lens probability rating.  A threshold of 66.7\% (two thirds the range of possible prediction scores) was chosen instead of a natural decision boundary of 50\% because it was desirable to further limit the number of false positives encountered during manual inspection. We chose to inspect 20,000 candidates because the objective of the rudimentary lens search was to demonstrate that our approach can also have practical applications in addition to comparing two model types. 20,000 images is a reasonable number to cover over a few work days of manual inspection. This amount of manual inspection is feasible and is similar to inspections done by \citep{jacobs19} where 16,729 candidates were manually inspected.
We present known lenses that the model was able to recover in Fig. \ref{fig:candidate1} and we present potential new  candidates that we were unable to find in previous works in Fig. \ref{fig:candidate2}. To search for originality of the new lens candidates we consulted the Masterlens Database of gravitational lenses \citep{masterlens}, the Simbad Astronomical Database \citep{simbad}, and the CASTLES Survey results \citep{castles}.

\begin{figure} % bar graph (original)
\centering

\includegraphics[scale=0.5]{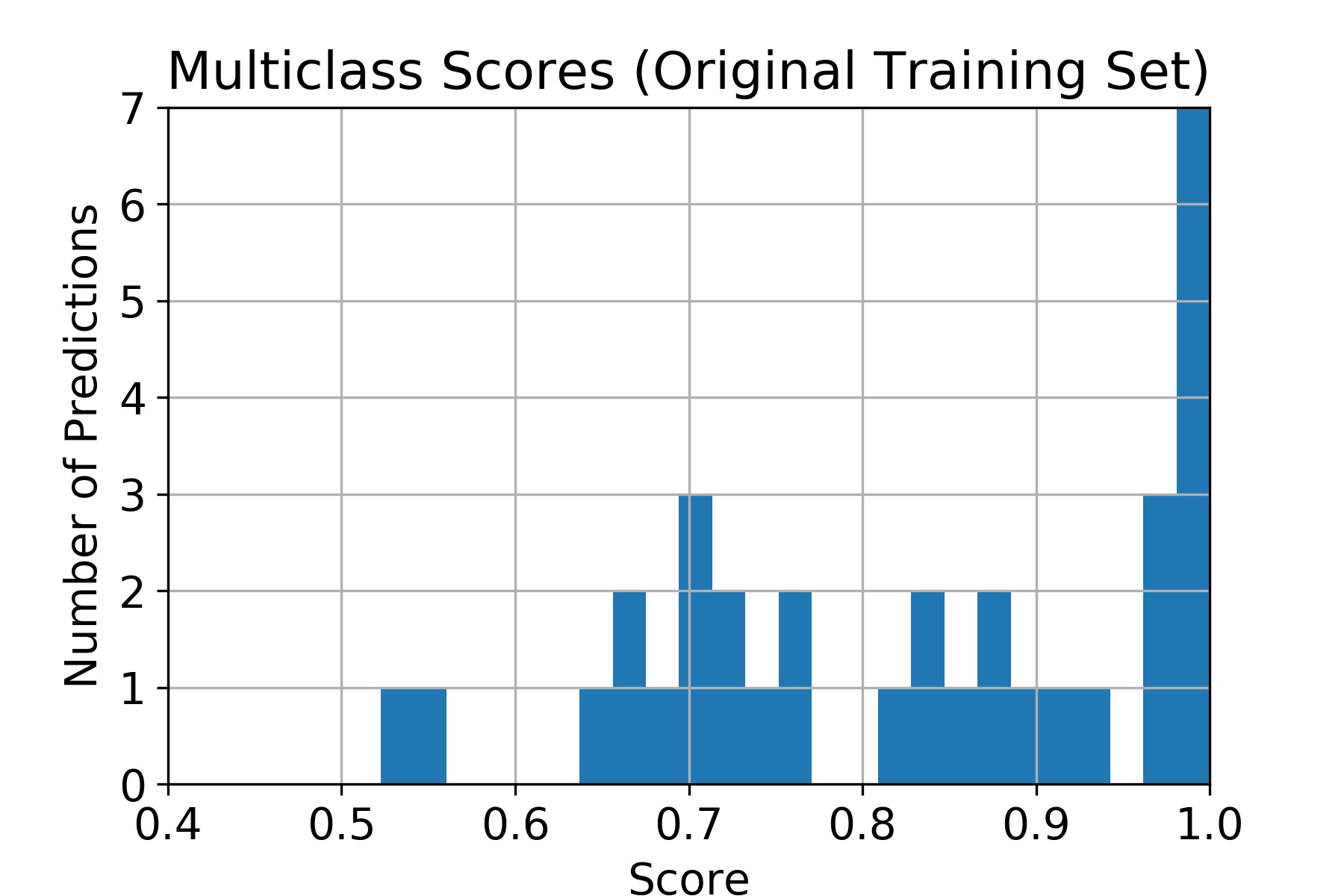}
\includegraphics[scale=0.5]{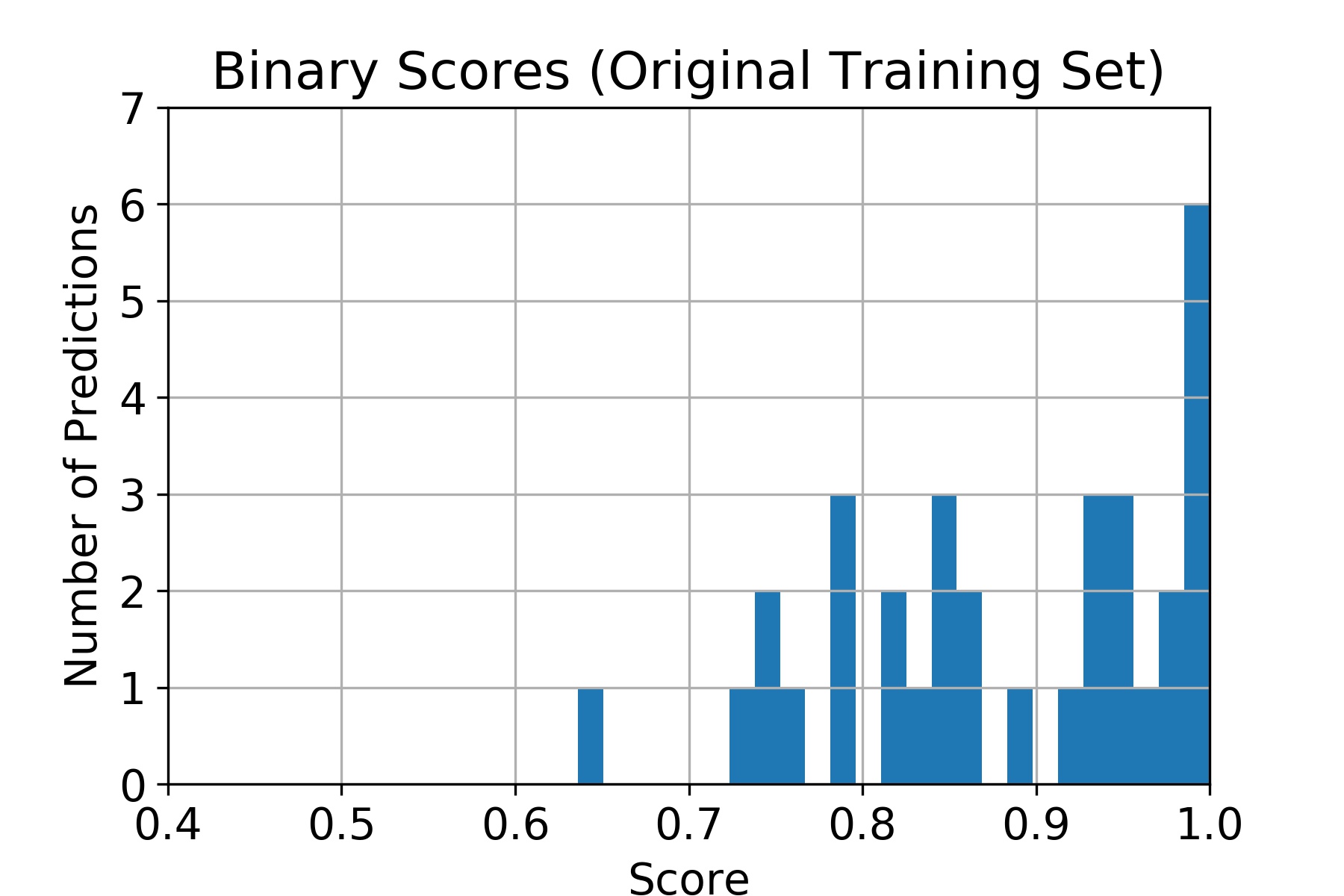}
\includegraphics[scale=0.5]{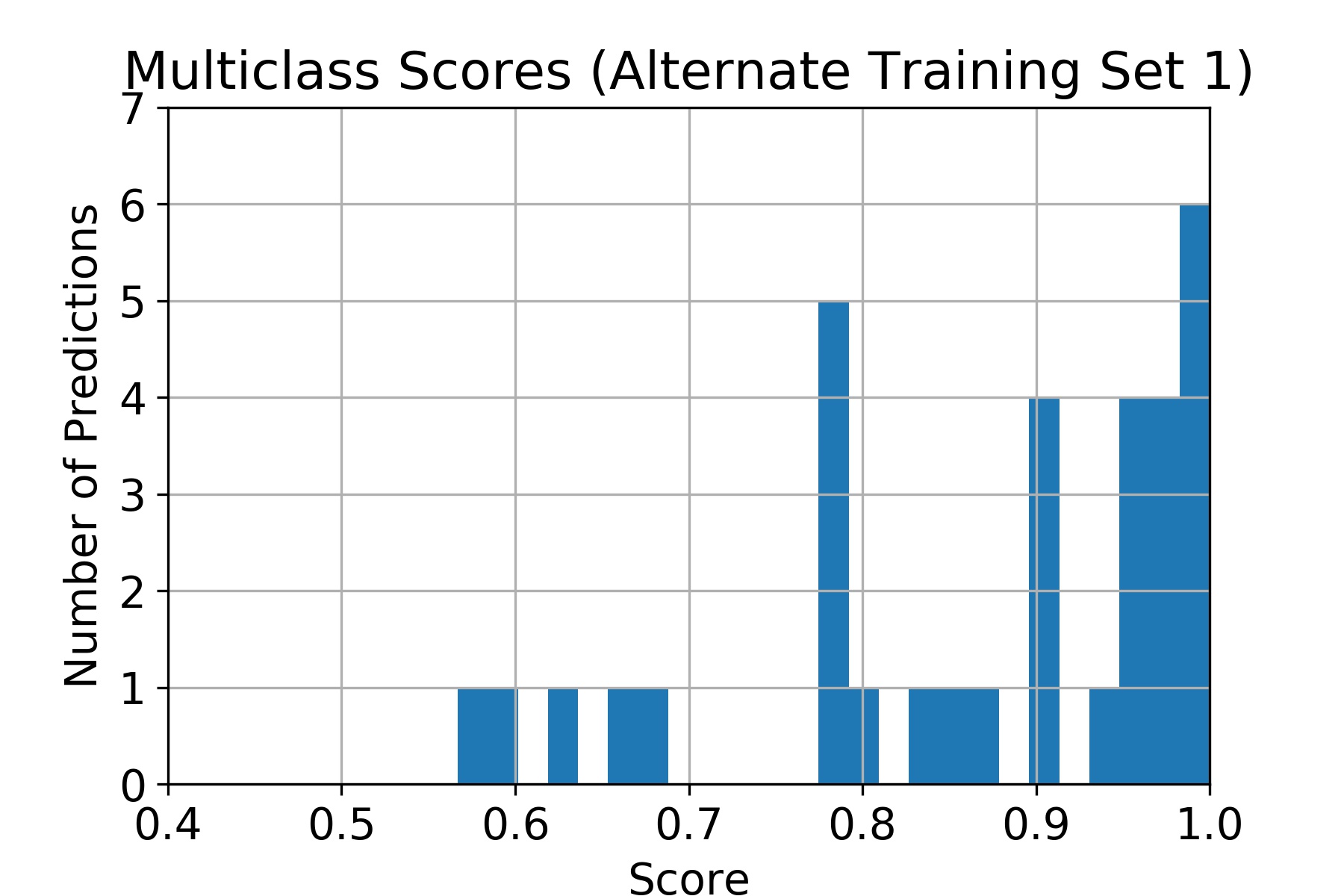}
\includegraphics[scale=0.5]{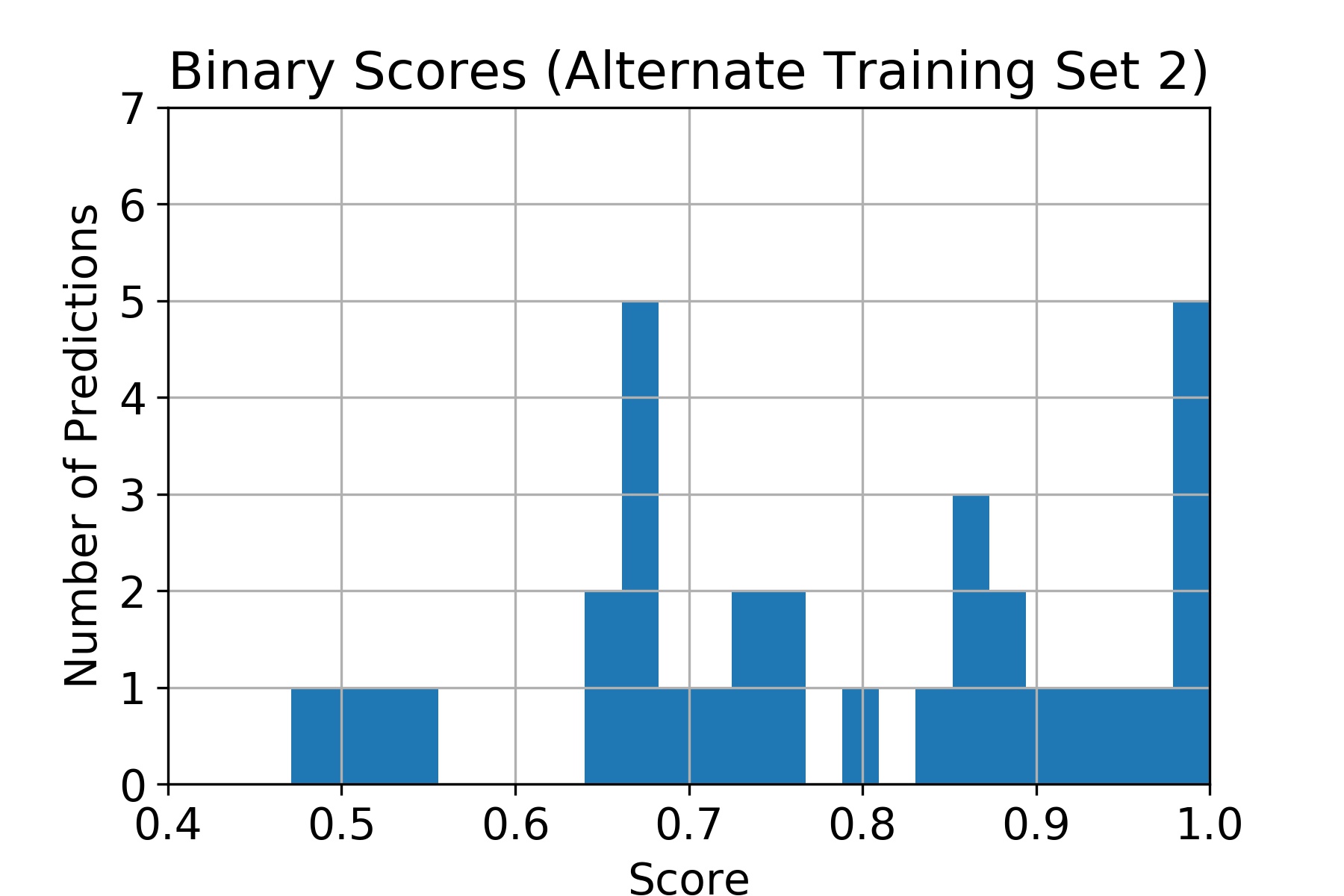}
\caption{Distribution of scores given to real lenses. Some scores are below 0.667 because only 14 of the 33 genuine lenses used were recovered from our HST search. The other 19 real lens we added afterwards specifically for this testing, but were not recovered during our HST lens search. }
\label{fig-histograms}
\end{figure}

Additionally, we examine the performance of the multi-class and binary models on real lens images. For this, we run the models on the 14 known lenses that were recovered from our search as well as an additional 19 lenses that we were able to find good quality HST images of in the F814W filter for a total of 33 (Using the same lens catalog as mentioned in the previous paragraph). We take the class with the highest prediction score as the model's labelling of each image, and calculate an average recall using ten models of each variety. The original multi-class model was able to recover 31 lenses on average , and the original binary model was able to recover 32 on average. The multi-class model trained using alternate training set 1 was able to recover 31 lenses, and the binary model trained with alternate training set 2 recovered 28 lenses. Table \ref{real-lens-table} shows the mean and median scores on the real lenses for each test case.  Figure \ref{fig-histograms} shows the distribution of scores in a histogram format for different cases. The number of false positives returned in each case must also be considered. Using the same 2500 negative test examples described in section 3.4, The original multi-class model returns 259 false positives, the original binary model returns 388 false positives. The multi-class model trained using alternate training set 1 returns 443 false positives, and the binary model trained with alternate training set 2 returns 199 false positives. These results are summarized in Table \ref{real-lens-table}. From these results, we can see that the binary model training on alternative dataset 2 returns the most pure samples, but has the poorest recall, while original binary model has the highest recall, but returns the most false positives. The two varieties of multiclass models have identical recall, but the original multiclass model has returns less false positives. Looking at the distribution of scores given to the real lenses, it can be seen that increasing the decision boundary from 66.7\% to a value over 75\%, for example, would not greatly reduce recall for most models. Increasing the decision boundary in that way might be a good strategy if a more pure sample is desirable.  Since the set of real lenses is very small, these results should not heavily factor into our comparison of the models. However, these results can at least further support that our models are applicable to real world problems.

%Table of metric results (original)
\begin{table}
  \centering
 
    \caption{Model Performances On Real Lenses}
    \begin{tabular}{ lllll }%|p{2.5cm}|p{2.5cm}|p{2.5cm}
\hline Model & Mean Real Lens Score & Median Real Lens Score & Real Lenses Recovered & False Positives \\
\hline
Multiclass (Original Training Set) & 0.82575 & 0.84266 & 31 & 259 \\
 Multiclass (Alternate Training Set 1) & 0.86935 & 0.91299 & 31 & 388 \\
Binary (Original Training Set)& 0.88077 & 0.89409 & 32 & 443\\
Binary (Alternate Training Set 2) & 0.781405 & 0.74836 &  28 & 199
%\hline
\label{real-lens-table}
\end{tabular}

\begin{tablenotes}
     \small
     \item Performance of the various models on real lenses. There are 33 real lenses and 2500 test negatives.

    \end{tablenotes}

%\caption{The error for each measurement is calculated at 95\% confidence}
\end{table}

\section{The discussion and the summary}
\label{sec:discusion}

We explored the possibility of using a multi-class classification scheme when searching for gravitational lenses with CNNs. First, we tested a four-class classification model and found that a three-class approach was a more suitable multi-class model because it resulted in less confusion between classes. Then we compared a three-class approach to the standard binary class approach and found that there were differences in some performance metrics. In our comparison of binary and multi-class  models, we found that we cannot conclusively say that a multi-class approach is more performant than the traditional binary approach, especially when considering the different training set splits that we investigated.

The number of classes and the type selected for a classification scheme may depend on the nature of the problem under study. For example, in an image-quality assessment, in which the goal is to separate 'usable' from 'unusable' images, a binary classification can constrain a model to find hidden or detailed patterns. This can happen because the unusable images can span a broad range of images in terms of quality. In other words, they vary greatly in quality and can have different character and nature. For example, an image taken in poor weather conditions can have a different pattern from one that has severe fluctuation problems in the background (Temmoorinia et al. 2019; submitted). Putting all 'unusable' images in a category may increase the level of confusion and force the model to produce a higher false-positive rate in the prediction. It is common practice in machine learning to use our knowledge about a data set to make an algorithm work better. However, in this paper, we are cannot conclude that the mutli-class approach is superior. It is worth noting that many factors affect the performance of a machine learning model not limited to the variety of classes that were chosen as part of the muticlass scheme, the quality of cutout-images for training, or even model architecture and optimization algorithm. We also show the results of applying our selected model to about 20 million cut-out images of size $100\times100$ , which are taken from  HST images (with the F814W filter and the ACS/WFC instrument).

%-------------------------------------------------------------------------------------------------------------------------

\section*{Acknowledgements}
We would like to acknowledge the help of Jon Willis for helpful discussion and insights.

CB acknowledges the support of a National Sciences and Engineering Research Council of Canada (NSERC) Graduate Scholarship.

\bibliographystyle{aasjournal}
\bibliography{bibl}

\end{document}